\documentclass[fleqn]{elsart}

\usepackage{amssymb}
\usepackage{amsmath}
\usepackage{amsfonts}
\usepackage{latexsym}

\newcommand{\version}{v3}

\newcommand{\beq}{\begin{equation}}
\newcommand{\eeq}{\end{equation}}
\newcommand{\beqa}{\begin{eqnarray}}
\newcommand{\eeqa}{\end{eqnarray}}

\newcommand{\bR}{\mathbb{R}}

\newcommand{\SM}   {Standard Model}

\begin{document}

\noindent  Physics Letters A 347 (2005) 8
\hfill     gr-qc/0503090 (\version)\vspace*{1cm}  

\begin{frontmatter}

\title{Coexisting vacua and effective gravity}

\author[a]{F. R. Klinkhamer},
\ead{frans.klinkhamer@physik.uni-karlsruhe.de}
\author[b,c]{G. E. Volovik}
\ead{volovik@boojum.hut.fi}
\address[a]{Institute for Theoretical Physics,
University of Karlsruhe (TH),\newline 76128 Karlsruhe, Germany}
\address[b]{Helsinki University of Technology,
Low Temperature Laboratory,\newline P.O. Box 2200, FIN--02015 HUT,  Finland}
\address[c]{Landau Institute for Theoretical Physics, 119334 Moscow, Russia}

\begin{abstract}
Our Universe may be a domain separated by physical phase
boundaries from other domain-Universes with different
vacuum energy density and matter content.
The coexistence of different quantum vacua is perhaps
regulated by the exchange of global fermionic charges
or by fermion zero modes on the phase boundary.
An example would be a static de-Sitter Universe
embedded in an asymptotically flat spacetime.
\end{abstract}


\date{24 May 2005} 
\end{frontmatter}

\section{Introduction}

Condensed-matter systems provide several examples of emerging quantum
fields (see  \cite{VolovikBook,Hu,FroggattNielsen} and references therein).
Effective gravity appears in the low-energy corner of
condensed matter either via the   metric  $g_{\mu\nu}({\bf r},t)$
or via the vierbeins $e_\mu^a ({\bf r},t)$.
These fields are functions of the coordinates ${\bf r}$
and $t$ of the underlying flat Galilei or Minkowski spacetime,
just as the other emerging fields (e.g., fermionic and  gauge
fields).

Emergent gravity does not require an \emph{a-priori} notion
of curved spacetime. The latter is a secondary  phenomenon
experienced by low-energy quasiparticles, whose dynamics in
the geometric-optics limit can be described in terms of motion
along the geodesic curves provided by the effective metric
$g_{\mu\nu}({\bf r},t)$. For the `poor physicist' of
Ref.~\cite{FroggattNielsen} who can only use quasiparticles with energies
low compared to the corresponding cut-off $E_{\rm  Planck}$,
the world is curved.

The most interesting condensed-matter models have effective
gravity emerging together with non-Abelian gauge fields
and Weyl fermions in systems with Fermi points
(generic points of level crossing at zero energy) \cite{VolovikBook}.
It may well be (though this has not yet been proven) that, if the
hierarchy of energy scales  is favorable, the dynamics of the effective
gravity is described by an Einstein--Hilbert action emerging at low
energy. For the `poor physicist' living in this particular type of
condensed matter,  any
connection to the `fundamental' underlying spacetime is  lost and
effective gravity is (at least, locally) indistinguishable from a
fundamental interaction.

These considerations force us to pay more attention
to approaches which consider the gravitational field as a conventional
non-Abelian spin--2 field in Minkowski spacetime \cite{VeltmanLesHouches1975}.
But, even if these theories in the low-energy corner
lead to the same local equations as Einstein gravity, the
global structure of the Universe can be different. The latter can, in
principle, be detected by following the particle trajectories
\cite{KlinkhamerVeltman}.

Moreover, the
mere existence of fundamental coordinates suggests
that in some cases
the so-called coordinate singularities, which appear
in certain solutions of the Einstein equations, are not gauge artifacts but
real physical singularities. This occurs if the  coordinate
transformation
needed to remove the singularity belongs to a class
of transformations forbidden in the global fundamental spacetime.

One example may be provided by a static Einstein universe
with positive cosmological constant
$\Lambda_0$ \cite{Einstein}, whose line element can be written as
\begin{equation}
ds^2=-dt^2 + \frac{dr^2}{1- r^2/R_0^2} +
r^2 \left( d\theta^2 + \sin^2\theta \,d\phi^2 \right)\,.
\label{EinsteinUniverse}
\end{equation}
Here, we have used relativistic units so that the velocity of light
in ``empty'' space has the value $c=1$ (later, we also set $\hbar=1$).
In the original Einstein theory, space is the three-dimensional (3D)
compact hypersurface of the 4D spacetime $\bR \times S^3$.
This Einstein Universe is geodesically complete and free from
singularities, which means that the coordinate singularity at $r=R_0$  is
a gauge artifact caused by the specific choice of coordinates on
the compact space $S^3$. More precisely,
the space described by the metric (\ref{EinsteinUniverse})
for $r \in [0,R_0)$ is only one hemisphere of $S^3$, whereas the whole
space of the original Einstein Universe has $r = R_0\,\sin\Psi$
for polar angle $\Psi \in [0,\pi]$;
cf. Chap. XII, Sec. 133 of  Ref.~\cite{Moller}.

Let us now consider the Universe (\ref{EinsteinUniverse})
from the point  of view of gravity as
a field  determined in  Minkowski spacetime
$M^4$. Since $M^4$ is infinite, the finite Einstein
Universe must be a part of the infinite system. One option is to replace
the coordinate singularity at $r=R_0$ with a real singularity
(or, better, a physical boundary between two regions) at  $r=R\leq R_0$.
The Einstein Universe lies inside the
boundary (interface) and the rest of Minkowski  spacetime outside:
\begin{subequations}\label{Einsteinbubble}
\begin{eqnarray}
ds^2 &=&
-dt^2 + \frac{dr^2}{1- r^2/R_0^2} +
r^2 \left( d\theta^2 + \sin^2\theta \,d\phi^2 \right)\,,
\quad r<R\leq R_0\,,
\label{EinsteinUniverse2}\\[2mm]
ds^2&=&
-dt^2 + dr^2 + r^2 \left( d\theta^2 + \sin^2\theta \,d\phi^2 \right)\,,
\quad r>R\,,
\label{Minkowski}
\end{eqnarray}
\end{subequations}
with constants $R$ and $R_0$ to be determined later.
Observe that this metric describes a static spacetime with an
\emph{inhomogeneous} spatial section: a Universe with
constant matter density and cosmological
constant $\Lambda=\Lambda_0 > 0$ inside, and an empty flat space with
$\Lambda=0$ outside. In contrast, the original Einstein Universe has a
homogeneous spatial section,
$S^3$ being a symmetric space. Indeed, the ``Universe'' from
Eqs.~(\ref{Einsteinbubble}ab) does have a \emph{center}
(i.e., a point with equal distance to the interface in each direction),
even though this may not be apparent to the `poor physicist' living
deep inside. For this observer, the inside region serves as a
static Einstein Universe.

The metric (\ref{EinsteinUniverse2})
obeys the Einstein equations, with matter density $\rho_{\rm m}$,
cosmological constant $\Lambda_0=\rho_{\rm vac}$, and
horizon  radius $R_0$ related by the following equations:
\begin{equation}
\Lambda_0=\rho_{\rm vac}= \half\, \rho_{\rm m} (1+3\,w)\,,\quad
\frac{1}{4\pi GR_0^2} =\rho_{\rm m}\,(1+w)\,,
\label{EinsteinSolution}
\end{equation}
where $P_{\rm m}=w\,\rho_{\rm m}$ is the  matter equation of state.
The total energy of the Einstein Universe including the interface
($r \leq R$) must be zero. This then matches the
Minkowski metric of the outer region ($r>R$), where matter is
absent and the cosmological constant is zero,
$\rho_{\rm m}=\rho_{\rm vac}=0$. The outer region exerts no pressure on
the inner region, which behaves as if there were no external environment.

However, the discontinuity of the metric  at $r=R$ and the jump
in the vacuum energy demonstrate that the singularity is
physical. The jump in vacuum energy requires a surface layer which separates
the ``true'' vacuum outside  ($\rho_{\rm vac}=0$ for $r>R$) from
the ``false'' vacuum inside ($\rho_{\rm vac}=\Lambda_0>0$ for $r<R$).
Within the surface layer, the original theory is modified.
In this approach, the
Einstein  Universe represents a bubble of false vacuum embedded in true
vacuum. Note also that the additional pressure from
the surface tension and the gravitational effects of the massive
boundary modify the balance between the outer and inner vacua.

Another example of a false-vacuum bubble is a de-Sitter Universe
\cite{deSitter} embedded in a space-time which is flat at
spatial infinity,
\begin{subequations}\label{deSitterbubble}
\begin{eqnarray}
ds^2 &=&-\left(1- r^2/R_0^2 \right)\, dt^2  +
\frac{dr^2}{1- r^2/R_0^2} +r^2 \, d^2\Omega\,,\quad r<R\leq R_0 \,,
\label{deSitterUniverse}\\
ds^2&=&-\left(1- 2GM/r \right)\, dt^2  +
\frac{dr^2}{1- 2GM/r} +r^2 \, d^2\Omega\,,\quad r>R \,,
\label{Schwa}
\end{eqnarray}
\end{subequations}
with solid angle element
$d^2\Omega \equiv d\theta^2 + \sin^2\theta \,d\phi^2$,
de-Sitter length scale \cite{Moller}
\begin{equation}
R_0 \equiv \sqrt{3/(8\pi G\Lambda_0)}\,,
\label{R0deSitterUniverse}
\end{equation}
and  constants $R$ and $GM$ to be determined later. Note that,
historically, Weyl \cite{Weyl} appears to have been the first to try
to replace the de-Sitter horizon at $r=R_0$ by a physical surface
layer 
and glue together two static de-Sitter Universes.

The embedding (\ref{deSitterbubble}ab) also has a
condensed-matter analogy; see Ref.~\cite{Laughlin} for a review.
This was originally discussed in relation to the event horizon of a
Schwarzschild black hole by Chapline \emph{et al.}
\cite{Chapline-etal}, and later by Mazur  and Mottola \cite{MazurCondensate}
and by Dymnikova and Galaktionov \cite{Dymnikova},
who viewed the event-horizon
coordinate singularity as a physical singularity. In their
treatment, the horizon is again a massive surface layer.
Similar junctions between different space-times have been considered
for inflation and wormhole  models
(see, e.g., Refs.~ \cite{BlauGuendelmanGuth,Visser}).

In both examples, a physical boundary separates two regions with
different values of $\Lambda$ and  serves as the interface between
two vacua with different values of
the vacuum energy. The problem of coexisting vacua in general
relativity and elementary particle physics is rapidly becoming
mainstream, not without the influence of condensed-matter physics.

Reference \cite{Vachaspati}, for example, considers  a
Friedmann-Robertson-Walker island embedded in a de-Sitter spacetime  with
nonzero cosmological constant. For elementary particle physics,
the so-called Multiple Point Principle  has been introduced
\cite{Nielsen-etal}, according to which Nature chooses the parameters of the
Standard Model so that two or more phases of quantum vacua have the same
energy density and can coexist. Coexistence of quantum vacua  has also been
discussed in relation to the cosmological constant problem  \cite{Coexistence},
where it was found  that the equilibrium value of $\Lambda$ in the exterior region
vanishes (based on an  analog from condensed-matter physics
and a thermodynamic argument applied to the
vacuum of relativistic quantum fields).

Furthermore, there may exist bubbles of false vacuum, whose stability
is provided by the conservation of a fermionic charge $Q$ trapped
within the bubble or on the interface.
These analogs of $Q$--balls \cite{Coleman} have been  discussed in
Ref.~\cite{lump} (see  also references therein).
Regarding the nature of the fermionic charge $Q$,
possible examples are the baryon charge $B$, the lepton charge $L$,
or the difference $B-L$, which is exactly conserved in the \SM~whereas
$B+L$ is not \cite{EWBNV}.

As a particularly simple case, we discuss in the next section
the embedded de-Sitter Universe and one possible type of  interface.
In the final section, we present some concluding remarks.

\section{De-Sitter bubble}
\label{DeSitterBubble}

\subsection{Pressure equilibrium}
\label{GeneralApproach}

Let us consider the de-Sitter--Schwarzschild  bubble from
Eqs. (\ref{deSitterbubble}ab) using thin-wall techniques
(see  \cite{BlauGuendelmanGuth,Visser} and references therein).
The interface (membrane)
at  radius $r=R$  could be a topological domain wall between two
almost-degenerate vacuum states. This wall is assumed to be very thin
compared to the characteristic size of the Universe, $\xi\ll R$, and  can be
considered as a delta-function singularity in the energy-momentum tensor
for the matter fields.

The detailed structure within the domain wall need not be
considered, since the general properties of the domain wall can be described
in terms of the (2+1)-dimensional energy-momentum tensor of the membrane,
which is characterized by two parameters, the surface energy density
$\epsilon$ and the surface tension $\sigma$.
In the thin-wall approach, the jump of the metric across the interface
can be expressed in terms of the surface energy density $\epsilon$
of the interface:
\begin{equation}
\epsilon= \frac{1}{4\pi GR}\,
\left[\,\sqrt{1-R^2/R_0^2}- \sqrt{1-2GM/R}\;\right]\,.
\label{dSSbubblesigma1}
\end{equation}
The equilibrium radius of the bubble can  be found from the thermodynamic
identity which expresses
the two-dimensional pressure $p$  of the interface (or the surface
tension $\sigma=-p$) through the surface energy,
\begin{equation}
p=-\sigma=-\frac{d( \epsilon A)}{dA}=
- \frac{1}{2R}\frac{d( \epsilon R^2)}{dR} \,,
\label{2Dpressure}
\end{equation}
where $A\equiv 4\pi R^2$ is  the area of the interface.
If  the energy  density $\epsilon$ does not depend on
$R$, one has simply $p=-\epsilon$.

Taking the derivative of Eq.~(\ref{dSSbubblesigma1})
    and using Eq.~(\ref{2Dpressure}), one obtains
\begin{equation}
p= \frac{1}{8\pi G R}\,\left[\,
\frac{1-GM/R}{\sqrt{1-2GM/R}}-
\frac{1-2R^2/R_0^2}{\sqrt{1-R^2/R_0^2}} \,\right]\,.
\label{dSSbubblesigma2}
\end{equation}
Provided the surface energy $4\pi R^2 \epsilon(R)$ is known,
Eqs. (\ref{dSSbubblesigma1})--(\ref{dSSbubblesigma2})
determine  the
radius $R$ and mass $M$ of the equilibrium bubble as functions of $R_0$
(or of $\Lambda_0 \equiv 3/8\pi GR_0^2$).

\subsection{Stabilization by fermion zero modes}
\label{FZM}

For the case of  fermionic charge concentrated at the interface,
the surface energy density $\epsilon$ clearly depends on $R$.
Let us suppose that the interface contains
fermion zero modes, that is,
massless (2+1)--dimensional relativistic  fermions with spectrum
$E(\mathbf{p})=c |\mathbf{p}|$. If  the chemical potential $\mu$ of
these fermions  is nonzero, the energy density of the interface includes
a contribution from the fermions,
\begin{equation}
\epsilon=\epsilon^0+   \frac{\mu^3}{6\pi}\,.
\label{SurfaceEnergyDensity}
\end{equation}
The surface tension is then given by
\begin{equation}
\sigma=-p=\epsilon-\mu \,n = \epsilon-\frac{\mu^3}{4\pi}=
\epsilon^0   -\frac{\mu^3}{12\pi}\,,
\label{SurfacePressure}
\end{equation}
where $n=\mu^2/4\pi$ is the number density of (2+1)--dimensional fermions.
(See Ref.~\cite{GuendelmanPortnoy} for a different mechanism to
stabilize the interface using (2+1)--dimensional gauge fields.)

The same surface tension can be obtained from Eq.~(\ref{2Dpressure})  if one
writes the  fermionic contribution to surface energy density in
Eq.~(\ref{SurfaceEnergyDensity}) in terms of the bubble radius for
a fixed number of fermions, $N=4\pi R^2 n =   R^2\mu^2\,$:
\begin{equation}
\epsilon(R)=\epsilon^0+   \frac{N^{3/2}}{6\pi R^3}\,,\quad
\sigma(R)=\epsilon^0-\frac{N^{3/2}}{12\pi R^3} \,.
\label{SurfaceEnergyDensity2}
\end{equation}
This dependence of the surface energy density allows us to regulate the
bubble  radius by changing the total fermionic charge $N$ of the interface.
With the fermion charge concentrated on the surface, this object
could perhaps be called a  $Q$--shell, but we will adhere to the more
general terminology of  $Q$--ball.

\subsection{Weakly gravitating $Q$--ball}
\label{WeaklyGravitatingBubble}

Let us first discuss the case  $R\ll R_0$. The  Universe then
occupies only a   small fraction
of the volume of the cosmological horizon and the effect of gravity  is
weak and  can be ignored.
In this limit,  we have a non-gravitating false-vacuum bubble
without matter. Its energy is given by
\begin{equation}
E=M =\frac{4\pi}{3} R^3 \,\Lambda_0+4\pi  R^2 \,\epsilon \,,
\label{ENonGravitUniverse(i)}
\end{equation}
which also follows from Eq.~(\ref{dSSbubblesigma1}) for  $R\ll R_0$.

Using the surface energy density from  Eq.~(\ref{SurfaceEnergyDensity2}), one
obtains the energy of
the de-Sitter bubble at fixed $N$:
\begin{equation}
E(R,N) =\frac{4\pi}{3} R^3 \,\Lambda_0+4\pi R^2\, \epsilon^0+
     \frac{2}{3R} \,N^{3/2}\,.
\label{ERN}
\end{equation}
The equilibrium condition $dE/dR=0$ corresponds to pressure balance
and gives the vacuum pressure inside the bubble,  $P_{\rm vac}\equiv
-\Lambda_0$,  in terms of the surface tension $\sigma=-p\,$:
\begin{equation}
P_{\rm vac}\equiv -\Lambda_0 = \frac{2\sigma(R)}{R} \,.
\label{PressureDeSitterNonG2}
\end{equation}
This last equation takes into account that the vacuum pressure outside
the bubble is zero, $P_{\rm vac}\equiv -\Lambda=0$. Nullification of the
cosmological constant in the exterior region may be a consequence of thermodynamic
stability for a system isolated from its environment  \cite{Coexistence}.

Equation (\ref{PressureDeSitterNonG2}) gives  the radius of the
$Q$--ball in terms
of $\Lambda_0$.  Since $\Lambda_0>0$, the  equilibrium false-vacuum $Q$--ball
exists only for  negative surface tension, which, according to
Eq.~(\ref{SurfacePressure}) or (\ref{SurfaceEnergyDensity2}), is
realized for large enough chemical potential $\mu$ of trapped fermions.
The condition for
ignoring gravity is $R\ll R_0$, which holds for $\sigma^2\ll \Lambda_0/G$.

For the corresponding weakly gravitating Einstein bubble
(\ref{Einsteinbubble}ab), the
radius $R$ is determined by a pressure-balance equation similar to
Eq.~(\ref{PressureDeSitterNonG2}). Now, the \emph{total} internal pressure
is compensated by surface tension:
\begin{equation}
P_{\rm vac}+P_{\rm m}\equiv- \Lambda_0+w\,\rho_{\rm m}=  \frac{2\sigma(R)}{R}\,,
\label{PressureEinsteinNonG}
\end{equation}
for matter equation of state  $P_{\rm m}=w\,\rho_{\rm m}$.

\subsection{Asymptotic horizons}
\label{ApproachingHorizon}

Let us return to the gravitating de-Sitter $Q$--ball ($Q$--shell) and
find the conditions under which $R$ can asymptotically approach
the  cosmological horizon $R_0$.
Analysis of Eqs.~(\ref{dSSbubblesigma1})--(\ref{dSSbubblesigma2}) 
shows that this occurs for negative surface tension
$\sigma$ under the following conditions\footnote{The
approximation of having surface energy density
$\epsilon$ and tension $\sigma$ as quantities independent of
gravity is only valid if the relative jump of the metric is
small. This gives the following refinement of conditions (\ref{conditions1}):
$\epsilon \ll  \Lambda_0/G|\sigma|\ll \sqrt{\Lambda_0/G} \ll |\sigma|
$.}:
\begin{equation}
\epsilon \ll   \sqrt{\Lambda_0/G} \ll |\sigma| \,.
\label{conditions1}
\end{equation}
Under these conditions, one has
\begin{equation}
1-2GM/R\approx 1- R^2/R_0^2\approx
\frac{3}{32 \pi}\, \frac{\Lambda_0}{G\sigma^2}\ll 1 \,.
\label{Approach}
\end{equation}

In the limit of large $|\sigma|$,  a cosmological horizon
is developing
asymptotically from the inside and a black-hole horizon from the outside.
For a `poor physicist'  living inside the bubble, the interior of
the bubble resembles  more and more a de-Sitter Universe, while, for
another `poor physicist' living outside,  the exterior of the bubble resembles
more and more a black hole.
And the smooth interface between the two vacua (with nonsingular metric)
starts to look like a genuine coordinate singularity.
In the same way,
the finite and curved Einstein Universe may be thought of as the
limiting case of a false-vacuum bubble (\ref{Einsteinbubble}ab)
in flat space.

However, the limiting case of a de-Sitter horizon cannot be realized
with an interface of
fermion zero modes as discussed in Sec. \ref{FZM}. The reason is that
Eqs.~(\ref{SurfaceEnergyDensity2}) and (\ref{conditions1}) require
the energy of the
interface without fermion-zero-mode contribution to be negative,
$\epsilon^0 < 0$. Hence, the static de-Sitter Universe with cosmological
horizon probably cannot be constructed using known interfaces, even
though suitable Planck-scale interfaces are not excluded.

\section{Discussion}

The static de-Sitter Universe \cite{deSitter}
(or the static Einstein Universe \cite{Einstein})
and the $Q$--ball with false vacuum \cite{Coleman,lump}
can be viewed as extreme limits of
the same physical object. For both, the vacuum energy is of crucial
importance, as well as the surface energy density and (negative) surface tension
of the membrane
separating the interior region from the external environment.
In addition, there may be an important role for a conserved
fermionic charge $Q$ inside the bubble or on the membrane.

In the limit of small surface energy density and large
negative surface tension, a cosmological horizon
is developing asymptotically from the inside and a black-hole horizon from
the outside. In this way, one nearly obtains a static de-Sitter Universe embedded in
a Schwarzschild spacetime as discussed previously
\cite{Chapline-etal,MazurCondensate,Dymnikova} for black holes with physical
horizons and false vacua in their interior. However,  the limiting
case  cannot be realized using known physical interfaces.

This also implies that, most likely,
emergent gravity is not able to incorporate the geodesically-complete
Einstein Universe with spatial section $S^3$.
(For a Hausdorff manifold, this would indeed be difficult to imagine
topologically.)
It, therefore, appears that the original static $S^3$ Einstein Universe
can exist only within the context of fundamental
general relativity.

\ack

F.R.K. thanks M.J.G. Veltman
for valuable discussions over the years.
The work of G.E.V. is supported in part
by the Russian Ministry of Education and
Science, through the Leading Scientific School grant $\#$2338.2003.2.
This work is also supported by the
European Science Foundation COSLAB Program.

\end{document}